# Some Explorations in Holomorphy[*]

MICHAEL DINE AND YURI SHIRMAN

*Santa Cruz Institute for Particle Physics*
*University of California, Santa Cruz, CA 95064*

**Abstract**

In supersymmetric theories, one can obtain striking results and insights by exploiting the fact that the superpotential and the gauge coupling function are holomorphic functions of the model parameters. The precise meaning of this holomorphy is subtle, and has been explained most clearly by Shifman and Vainshtein, who have stressed the role of the Wilsonian effective action. In this note, we elaborate on the Shifman-Vainshtein program, applying it to examples in grand unification, supersymmetric QCD and string theory. We stress that among the "model parameters" are the cutoffs used to define the Wilsonian action itself, and that generically these must be defined in a field-dependent manner to obtain holomorphic results.

Submitted to *Physical Review D*

[*] Work supported in part by the U.S. Department of Energy.

## 1. Introduction

It is possible to make very powerful statements about four-dimensional supersymmetric theories using some minimal information about conventional global and local symmetries, combined with the constraints that supersymmetry implies on the effective action. These techniques have been used to explore the nature of dynamical supersymmetry breaking[1,2] and to prove powerful non-renormalization theorems in string theory in only a few lines.[3,4] More recently, the idea that the effective superpotential should be an analytic function of the parameters has given new insight into the non-renormalization theorems of supersymmetric field theories, shedding light on the non-perturbative behavior of these theories, even in their strong coupling regimes.[5,6] It has also been used to consider properties of non-perturbative string theory.[7,8] All such arguments rely on the fact that the effective low energy lagrangian is specified by three functions, two of which are holomorphic functions of the chiral fields: the superpotential, $W$, and the gauge coupling function, $f$.

Yet there is a cloud which hangs over the use of arguments of this type. If one examines perturbation theory, one finds that these functions appear to obtain non-holomorphic corrections in low orders in theories with massless particles. It was Shifman and Vainshtein who explained that the problem is to differentiate between a "Wilsonian action," in which states with mass or momentum above some value have been integrated out, and a more conventional effective action.[9] Their arguments also resolved a set of paradoxes connected with the "multiplet of anomalies." Still, it is often unclear how to implement these ideas in practice, and there is great unease about the consequences of holmorphy.

In this note, we elaborate the Shifman-Vainshtein (SV) program. Fol-lowing ref. 5, we view the parameters of a supersymmetric theory as vev's of chiral fields. In string theory, this is generally the case. In field theories, this is a powerful device to constrain the possible dynamics.[5,6] However, field theories (including the *Wilsonian* effective actions which describes string models at low energy) contain parameters which do not appear explicitly in the lagrangian: the cutoffs. If these cutoffs are not chosen properly, one can induce non-holomorphicity; in particular, field-dependent redefinitions of these cutoffs lead to (in general non-holomorphic) field-dependence in the action. This viewpoint leads us to rephrase the SV program in terms of field (or parameter)-dependent cutoffs. Two types of non-holomorphicity have been discussed in the literature. First, SV have pointed out that, quite generally, at two loops and beyond, the gauge function $f$ is not holomorphic as a function of the coupling constants. Second, Dixon, Kaplunovsky and Louis (DKL)[10] have noted that in models in which there are massless states and in which the mass matrix has a non-trivial field dependence (e.g. on some moduli fields, as in string theory), there is generically some non-analyticity already at one loop. We will understand, in fact, both classes of problem in terms of field-dependent cutoffs.

The basic problem, and the resolution we will describe, are easy to understand. The problem has two aspects. First, why is it crucial to deal with a Wilsonian action? In theories with massless fields, the conventional "one-particle irreducible action" is not local. It contains, for example, at the loop level terms involving $\log(p^2)$. As a result, this action cannot necessarily be written according to the standard rules in terms of a superpotential, Kahler potential, and gauge coupling function. The appearance of non-holomorphic functions of the chiral fields, much less of the parameters, in the non-local action should not be a surprise. The Wilsonian action, defined by integrating over momenta above some cutoff, on the other hand, is necessarily local, and, provided the regulatory preserves supersymmetry, must be expressible in the standard supersymmetric form. It thus involves a superpotential and a gauge coupling function which must be holomorphic functions of any chiral fields. For models where one can add explicit mass terms for fields, this has been verified through two loops in ref. 11. In this note, we will illustrate this point with a number of additional examples.

Even in the context of the Wilsonian action, however, there are additional issues which must be faced in understanding holomorphy. The basic argument for analyticity is that the terms in the superpotential can be viewed as vev's of chiral fields. More precisely, we can view changes in these parameters as arising due to changes in the vev's of chiral fields. For example, we might write some mass parameter as $m_o + f(m_o)\phi$, where $<\phi>= 0$. By our argument above, the superpotential and gauge coupling function are necessarily holomorphic functions of $\phi$. But their dependence on $m_o$ is more subtle. It is clear that in order that quantities be analytic in $m_o$, we must parameterize the fields sensibly. If we redefine, for example, the chiral fields by non-holomorphic functions of the parameters, we will obtain non-holomorphic expressions. This problem already exists at tree level, as we will illustrate in the model of DKL. At the loop level, there is an additional difficulty. The Wilsonian action itself contains parameters not explicitly present in the lagrangian – the cutoff(s) used to define it. Changing these cutoffs by field- or coupling-dependent amounts changes the couplings by field or coupling-dependent amounts. As a result, only for a special choice of cutoffs do we expect to obtain results analytic in the parameters. In many cases, these choices are equivalent to non-polynomial redefinitions of the parameters. We will see, however, when we consider grand unified theories, that this description is not always suitable, and so we prefer the cutoff language.

In some cases, as we will note (by a modest extension of the Shifman-Vainshtein discussion), one can choose cutoffs so that everything is guaranteed to be analytic from the start. However, one can (at least at low orders of perturbation theory) give many definitions of the Wilsonian action, and not all of these give manifestly holomorphic results. We will consider two schemes, which are variants of the usual minimal subtraction and momentum schemes, and see how the cutoffs must be redefined in order to obtain holomorphic results. The lessons we will draw from all of this are simple. In trying to infer the consequences of holomorphy in a given situation, one must be careful about field redefinitions and allow for the possibility of non-holomorphicity arising from field-dependent cutoffs. We will turn to an examination of how considerations of this kind apply to some of the appliciations of holomorphy mentioned above. In particular, we will consider supersymmetric QCD, with various numbers of flavors and colors. For $N_f < N$, it is possible to compute the form of the effective superpotential. The dynamical calculation is different in different cases; we will verify that in all cases this superpotential is holomorphic in the appropriate variables. This is consistent with the remarkable arguments of refs. 6 and 12, which permit one to perform computations in what would seem to be inappropriate limits.

Finally, we will discuss how our considerations extend to string theory. In string models with low energy supersymmetry, all of the parameters are determined by expectation values of chiral fields. However, only the Wilsonian action is guaranteed to be expressible as a holomorphic function of these fields.[10] Based on our field theory experience, we will discuss procedures for defining the Wilsonian action, and the problem of making a suitable choice of scale. This is important to understanding constraints on non-perturbative effects following from symmetries.[7,8]

It is important to stress again that the discussion of this note is simply an elaboration on the ideas of Shifman and Vainshtein. Hopefully, it will be of value to those trying to understand how these considerations apply in various contexts. All of our considerations will be in the context of global supersymmetry. Important additional considerations which arise in the context of local supersymmetry have been discussed recently by Kaplunovsky and Louis.[13]

## 2. Holomorphy at Tree Level

Before jumping into loop computations, it is instructive to examine how the problem of holomorphy appears at tree level. As an example, we consider a model due to DKL.[10] This model was constructed to reproduce certain features of one loop string computations. We will consider the model at one loop in the next section, but already at tree level it contains some subtle features. The model is based on the gauge group $E_6$ (this is not essential; indeed, at tree level, the gauge interactions will be irrelevant). There are two 27's, $27_1$ and $27_2$, and one $\overline{27}$. There are two singlet fields, $\phi_1$ and $\phi_2$ (to be thought of as moduli). The superpotential is taken to be

$$W_{dkl} = \phi_1 27_1 \overline{27} + \phi_2 27_2 \overline{27} \qquad (2.1)$$

We want to explore the analyticity properties of this model as a function of $\phi_1$ and $\phi_2$. At a generic point in the "moduli space," the $\overline{27}$ pairs with a linear combination of the 27's and gains mass.

At tree level, we would like to integrate out the massive field, and obtain an effective lagrangian for the light fields. As the model stands, this is rather trivial, since the superpotential of the light field vanishes. However, if we add to the original superpotential cubic (and possibly higher order) terms, the problem becomes more interesting. For example, take

$$W_{int} = \lambda_1 27_1^3 + \lambda_2 \overline{27}^3. \qquad (2.2)$$

It is helpful to organize the computation by writing

$$\phi_1 = m_1 + \delta\phi_1 \qquad \phi_2 = m_2 + \delta\phi_2 \qquad (2.3)$$

where now $\phi_1$ and $\phi_2$ have no vacuum expectation values. We can now write the massive field as

$$h = \frac{m_1 27_1 + m_2 27_2}{\sqrt{|m_1|^2 + |m_2|^2}}. \qquad (2.4)$$

This field has mass $m_h = \sqrt{|m_1|^2 + |m_2|^2}$. The light field is the orthogonal linear combination. Taking $m_1$ and $m_2$ as complex, we are led to define

$$l = \frac{-m_2^* 27_1 - m_1^* 27_2}{\sqrt{|m_1|^2 + |m_2|^2}}. \qquad (2.5)$$

(With this choice, $27_1^\dagger 27_1 + 27_2^\dagger 27_2 = h^\dagger h + l^\dagger l$.)

Inverting these relations gives

$$27_1 = \frac{m_1^* h - m_2 l}{\sqrt{|m_1|^2 + |m_2|^2}}$$

$$27_2 = \frac{m_2^* h + m_1 l}{\sqrt{|m_1|^2 + |m_2|^2}}$$

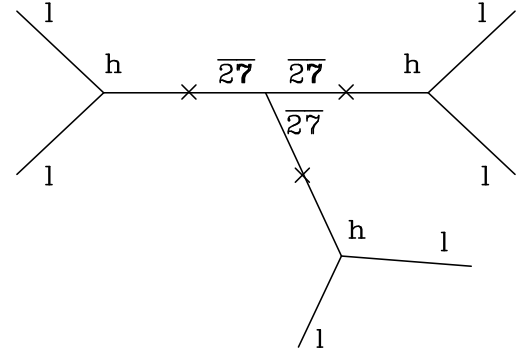

**Fig. 1.** Diagram contributing to an $l^6$ coupling in the effective superpotential for the light fields.

Now we can try to integrate out the massive field. The leading effect is just to rewrite $\lambda_1 27_1^3$ in terms of $l$. It is clear that this term will only be analytic in $m_1$ and $m_2$ if we rescale the field $l$, $l \to \frac{l}{\sqrt{|m_1|^2+|m_2|^2}}$. The leading term in the superpotential is then just

$$W_{eff} = -\lambda_1 (m_2 l)^3. \qquad (2.6)$$

The Kahler potential, however, is now, to this order

$$K = (|m_1|^2 + |m_2|^2) l^\dagger l. \qquad (2.7)$$

Integrating out the massive field gives terms which are higher order in $l$. The leading correction is proportional to $l^6$. It is given by the Feynman diagram of fig. 1 written in superspace. It is manifestly proportional to $\int d^2\theta$, and thus might be interpreted as a term in the superpotential. However, it is

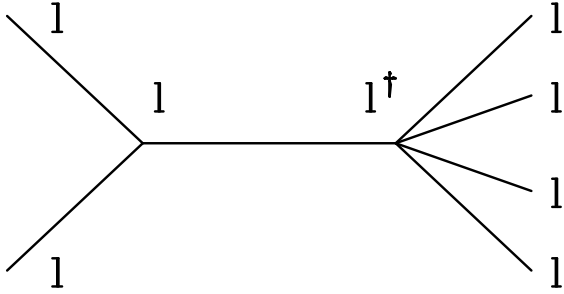

**Fig. 2.** Combining the corrections to the Kahler potential with the lowest order superpotential to yield the $l^6$ coupling.

proportional to
$$\lambda_1^3 \lambda_2 m_2^6 \frac{m_1^{*3}}{\sqrt{|m_1|^2+|m_2|^2}^6} l^6 \tag{2.8}$$

where $l$ is the *rescaled* field. This is clearly not holomorphic. This is all the more puzzling since the superpotential is manifestly an analytic function of $\phi_1$ and $\phi_2$.

The resolution to this puzzle is simple. The amplitude can be reproduced in the low energy theory provided we add to the Kahler potential a term of the form
$$\delta K = \frac{m_1^{*3}}{\sqrt{|m_1|^2+|m_2|^2}^6} \lambda_1^2 \lambda_2 l^\dagger l^4 \tag{2.9}$$

Then the diagram of fig. 2 reproduces the coupling above. One can verify that this procedure is correct by including fluctutations in the moduli. For example, replacing $m_1^* \to m_1^* + \phi_1^\dagger$, we expect terms in the Kahler potential such as
$$\frac{m_1^{*2}\phi_1^\dagger}{\sqrt{|m_1|^2+|m_2|^2}^6} \lambda_1^2 \lambda_2 l^\dagger l^4. \tag{2.10}$$

It is straightforward to check that these are present with the correct coefficients.

So we have encountered one of the problems described above. It is clearly necessary to carefully define the fields. It is also necessary to be careful how one organizes the effective lagrangian into Kahler potential and superpotential. In particular, iterating the Kahler potential and superpotential can lead to couplings which, in terms of fluctuating fields, have the structure of a superpotential. Indeed, R. Leigh suggests another approach which makes clear that the tree level effective superpotential can be written as a holomorphic function.[14] Take as the heavy field,
$$h = m_1 27_1 + m_2 27_2.$$

Take as an interpolating field for the light field,
$$l' = -m_2 27_1 + m_1 27_2.$$

These fields are no longer orthogonal (e.g. they have mixed kinetic terms). Still, one can integrate out the field $h$ (and $\overline{27}$) at tree level by solving its classical equations of motion. The result is holomorphic effective action for $l'$ which will reproduce the $S$ matrix for the light field.

## 3. The DKL Model at One Loop: A First Encounter With Field-Dependent Regulators

We now turn to the analysis of this model at one loop, and the problem raised by DKL. If one computes the gauge coupling function in a theory of this kind, one obtains a result proportional to the logarithm of the mass of the heavy field, i.e. $\ln(|m_1 + \delta\phi_1|^2 + |m_2 + \delta\phi_2|^2)$. To understand how this non-holomorphicity results, we can modify the theory slightly in a way which permits a simple definition of a Wilsonian action. Add to the theory an additional $\overline{27}$, which does not appear in the superpotential; call this field $\overline{27}_2$. Obviously this theory suffers from the same difficulty as the original one. However, now we can define a Wilsonian effective action by adding a mass term of the form

$$M 27_1 \overline{27}_2, \tag{3.1}$$

where we will be interested in $M \ll m_1, m_2$. This has the effect, as noted in ref. 11, of cutting off momenta $p \ll M$. It is easy to check that for this Wilsonian action, the coupling function is holomorphic; it is proportional to $\ln(M(m_1 + \delta\phi_1))$.

Let us examine the structure of this modified theory more closely. Introducing the Wilsonian action in this way, we write a general amplitude as

$$\Gamma = \Gamma_W + (\Gamma - \Gamma_W) \tag{3.2}$$

where $\Gamma_W$ denotes the amplitude one would compute by taking matrix elements of the Wilsonian action. This is an instruction to take the low energy action for the gauge fields as the action one would obtain by integrating out a massive field, and to regulate the theory with a Pauli-Villars mass. It is easy to see that the physical mass of what we earlier called the field $l$, with this mass term, is

$$M \sqrt{\frac{|m_1 + \delta\phi_1|^2}{|m_1 + \delta\phi_1|^2 + |m_2 + \delta\phi_2|^2}}. \tag{3.3}$$

In other words, the regulator mass is *field-dependent*. Indeed, it is clear that loops computed in the low energy theory with this field-dependent cutoff reproduce the non-analyticity observed above in the gauge coupling function.

Having seen that field-dependent regulators are an inevitable aspect of the holomorphy problem, we proceed to consider some more general examples.

## 4. Gauge Theories: Three Regulators

In this section, we restate in a slightly different language the results of SV. Consider, first, a $U(1)$ gauge theory, with two massless chiral fields, $e$ and $\bar{e}$. In such a theory, the function $f$, computed at some scale, should be a holomorphic function of the gauge coupling and a "$\theta$-angle." In other words, think of the gauge coupling, $\frac{1}{g^2}$, as the vev of a chiral field,

$$S = \frac{1}{g^2} + ia + \ldots \tag{4.1}$$

$f$ should then be a holomorphic function of $S$. However, the theory is invariant under shifts of $a$, so the only allowed terms in $f$ are

$$f = -\frac{1}{4}S + b$$

where $b$ is independent of $S$ ($g$). This would suggest that there is no correction to the $\beta$-function beyond one loop, contradicting well-known results. SV provided the solution to this paradox: the coupling for which analticity holds (which they referred to as the Wilsonian coupling) is related to a more conventional one by a non-polynomial redefinition of the coupling.

We can restate their arguments slightly in a way which makes this conclusion obvious. For a theory such as this one, we can define a Wilsonian action, $S_W(m_o)$ as the action for a theory in which we add a mass term, $m_o e \bar{e}$ to the superpotential. This has the effect of eliminating momenta smaller than $m_0$. With this rule, we break the full amplitude for some process, $\Gamma$, into a two pieces, similar to those of our previous example:

$$\Gamma = \Gamma_W(m_o) + (\Gamma - \Gamma_W(m_o)). \tag{4.2}$$

where $\Gamma_W(m_0)$ is the amplitude computed with the Wilsonian action. This rather trivial decomposition is just the instruction to compute in the low energy theory with the Wilsonian action and a Pauli-Villars regulator. More precisely, we should introduce two masses, $m_o$ and $m'_o$, $m_o \gg m'_o$, and define the Wilsonian action as the difference of the action computed with these two masses.[*] Because the theory with this regulator is completely finite, and because the action is necessarily an analytic function of all of its parameters, one expects that the action, written as a function of $m'_o$, $m_o$, and $S$ is completely analytic, and this is indeed the case. In particular, written as a function of these quantities, the gauge field terms in $S_W$ are corrected only at one loop:

$$S_W = -\frac{1}{4} S + b_o \ln(m'_o/m_o) \tag{4.3}$$

where $b_o$ is the first term in the usual $\beta$-function.

---

[*] As an intermediate step, one can use dimensional regularization (reduction) to regulate the separate actions.

However, the masses $m_o$ and $m'_o$ are themselves renormalized. At one loop order, denoting the renormalized parameters by $m$ and $m'$

$$\frac{m'}{m} = Z \frac{m'_o}{m_o} = \frac{m'_o}{m_o} \left( \frac{g(m)}{g(m')} \right)^{-c/b_o}. \tag{4.4}$$

$c \frac{g^2}{16\pi^2}$ is the anomalous dimension of the charged fields. This connection of the mass renormalization and the anomalous dimensions, as is well known, follows from the fact that there is no renormalization of the superpotential, so the only mass renormalization arises from wave function renormalization. A straightforward one loop calculation gives $c = 4$.

As an aside, we note that the calculation of this anomalous dimension has a few amusing aspects. First, if one works in a manifestly supersymmetric fashion, in terms of supergraphs, the statement that any fermion mass is only renormalized as a result of wave function renormalization means that the wave function renormalization must be gauge invariant. This indeed turns out to be the case. In addition, if one uses the standard supergraph rules, there are two diagrams; each is infrared divergent, due to terms proportional to $1/k^4$ in the propagators. These divergences cancel between the two diagrams which contribute, leaving $c = 4$.

If we rewrite $S_W$ in terms of the physical, renormalized scales, using eqn. (4.4), we discover the dependence on scales expected from the two-loop $\beta$-function:

$$\frac{8\pi^2}{g(m')^2} = \frac{8\pi^2}{g(m)^2} + b_o \ln(m'/m) + \frac{b_1}{b_o} \ln\left(\frac{g(m)}{g(m')}\right) + \mathcal{O}(g^2) \tag{4.5}$$

where

$$b_1 = -8. \tag{4.6}$$

In other words, written in terms of the "bare" cutoff masses, the $\beta$-function is renormalized only at one loop. However, written in terms of the physical, renormalized cutoffs, we recover a conventional $\beta$-function. As pointed out by SV, the validity of eqn. (4.3) means that there is an exact relation in this theory between the $\beta$-function and the anomalous dimension to all orders of perturbation theory. Note that as usual, a change in the scale corresponds to a redefinition of the coupling; in the present case, the coupling in terms of which the action is analytic,

$$\frac{8\pi^2}{g_a^2} = \frac{8\pi^2}{g^2} + \frac{b_1}{b_o} \ln(g) \tag{4.7}$$

.

What if we use some other sort of regulator? We can, for example, calculate the coupling constant renormalization using dimensional regularization[*] and the $\overline{MS}$ scheme. In this calculation, this regulator cuts off momenta, $\Lambda$, larger than

$$\Lambda = \mu \epsilon^{-1/\epsilon}. \tag{4.8}$$

So we can again define a *Wilsonian* action as the difference of two regulated actions, with scales $\mu$ and $\mu'$.[†] In this formulation, we will, at two loops, obtain a conventional-looking result for the action, with coupling

$$\frac{8\pi^2}{g(\mu')^2} = \frac{8\pi^2}{g(\mu)^2} + b_o \ln(\mu'/\mu) + \frac{b_1}{b_o} \ln\left(\frac{g(\mu)}{g(\mu')}\right) + \mathcal{O}(g^2) \tag{4.9}$$

It is clear from our previous discussion what has happened. In order to define the Wilsonian action, it is necessary to introduce cutoff parameters. Only for

$$\frac{\mu'}{\mu} = \frac{\mu'_o}{\mu_o} \left(\frac{g(\mu)}{g(\mu')}\right)^{-c/b_o}. \tag{4.10}$$

---

[*] More precisely dimensional reduction. Note we are content to use in this analysis regularization schemes which work only to low orders.

[†] A dimensionally regulated action is not, in general, a Wilsonian action, since dimensional regularization does not really act as a cutoff in integrals with power divergences. Similar issues arise with the momentum space scheme discussed below. We thank Joe Polchinski for a discussion of this issue.
suitable definitions of these parameters, will the action be holomorphic. Here it is necessary to define

The action as a function of these new parameters, $\mu_o$ and $\mu'_o$, is also an analytic function of $S$.

There is still a third regulator which is convenient for discussing the Wilsonian action: a momentum space regularization scheme. We can simply define the Wilsonian action by specifying the values of certain Green's functions at a suitable Euclidean momentum point, $M$. This, again, has the effect of cutting off momenta below the scale $M$. Actually, we want to take, again, the difference of two such regulated actions, with scales $M'$ and $M$. This gives an expression for the action similar to that of equation (4.9). Again, one can define new masses as in eqn. (4.10) such that the action is analytic.

These last two regulators are convenient for discussing non-abelian theories and chiral theories. Here there is no convenient Pauli-Villars type regulator available, but it is clear that what we want to do is define suitable rescaled cutoffs so as to eliminate the two (and higher) loop renormalizations of the couplings. There is no obstacle to doing this. Indeed, the equations are identical to those we have discussed above for the momentum space regulator or $\overline{MS}$ regulator, provided the $\beta$-functions are simply taken appropriately.

Before considering non-perturbative questions, let us apply these ideas to an $SU(5)$ GUT. To make the equations simple, consider a theory with an adjoint, $\Sigma$, of chiral fields, but with no other matter fields. In terms of the bare fields, write the lagrangian as

$$\mathcal{L}_{gut} = \int d^4\theta Z_A^{-1} \operatorname{tr} \Sigma_o^\dagger \Sigma_o + \int d^2\theta \operatorname{tr} m_o \Sigma_o^2 + \operatorname{tr} \frac{\lambda_o}{3} \Sigma_o^3. \tag{4.11}$$

Here $Z_A$, the wave function renormalization factor, is given by

$$Z_A = \left(\frac{g(M)}{g(M_V)}\right)^{-2}. \qquad (4.12)$$

This lagrangian has a minimum which breaks $SU(5)$ to $SU(3) \times SU(2) \times U(1)$,[15]

$$\Sigma_o = \sigma_o \, diag(2,2,2,-3,-3) \qquad (4.13)$$

where

$$\sigma_o = 2\frac{m_o}{\lambda_o}. \qquad (4.14)$$

In this vacuum, the vector masses go as

$$M_V = 5\sqrt{2}g(M_V)\sigma = g(M_V)\sigma_o/\sqrt{Z}. \qquad (4.15)$$

The remaining members of the adjoint have mass of order $m = Zm_o$; the octet, triplet and singlet have masses $5/2m$, $5/2m$, and $1/2m$, respectively.

We would like to consider the Wilsonian effective action obtained by integrating from a scale, $M$, well above the GUT scale to a scale, $\mu$, well below the GUT scale. Using the conventional renormalization group analysis, we can determine the coupling constant at the scale $\mu$ ($\mu \ll M_V$) for each group:

$$\frac{8\pi^2}{g^{(i)2}(\mu)} = \frac{8\pi^2}{g^2(M)} + \tilde{b}_o^{(i)} \ln(\mu/M) + \frac{\tilde{b}_1^{(i)}}{\tilde{b}_o^{(i)}} \ln\left(\frac{g(M)}{g^{(i)}(\mu)}\right)$$
$$+ (b_o - \tilde{b}_o^{(i)} + \tilde{N}^{(i)}) \ln(M_V/M) - \tilde{N}^{(i)} \ln(m/M) + \left(\frac{b_1}{b_o} - \frac{\tilde{b}_1^{(i)}}{\tilde{b}_o^{(i)}}\right) \ln\left(\frac{g(M)}{g(M_V)}\right). \qquad (4.16)$$

In this expression, $b_0 = 2N = 10$ and $b_1 = 0$ denote the one and two loop $\beta$-function coefficients of the high energy theory; $\tilde{b}_o^{(i)}$ and $\tilde{b}_1^{(i)}$ denote the corresponding quantities for the three low energy groups, and $\tilde{N}^{(i)}$ is the Casimir of the adjoint representation (associated with the massive octet, triplet and singlet). Note that for the low energy $U(1)$ these quantities vanish. The various terms here can be inferred simply by noting that above the scale $M_V$, the couplings flow with the $\beta$-function of the high energy theory. For the massive fields, we can then simply replace $\mu$ with the appropriate threshold. The thresholds for the vector fields are at $M_V$ and those for the octet, triplet, and singlet are at $m$ (up to a constant of order one).

Now rewriting $M_V$ and $m$ in terms of bare quantities, using equation (4.12) and the explicit forms of the $\beta$ functions, we obtain:

$$\frac{8\pi^2}{g^{(i)2}(\mu)} = \frac{8\pi^2}{g^2(M)} + \tilde{b}_o^{(i)} \ln(\mu/M) + \frac{\tilde{b}_1^{(i)}}{\tilde{b}_o^{(i)}} \ln\left(\frac{g(M)}{g^{(i)}(\mu)}\right)$$
$$+ (2N - 2\tilde{N}^{(i)}) \ln g(M) + (2N - 2\tilde{N}^{(i)}) \ln(\sigma_o/M) - \tilde{N}^{(i)} \ln(m_o/M). \qquad (4.17)$$

These expressions are analytic in the bare parameters, $m_o$ and $\lambda_o$. However (apart from the $U(1)$, which does not involve the scale $\mu$ at all) they are only analytic in $g(M)$ if we define independent parameters, $\mu^{(i)}$, for each gauge group, i.e. we let $\mu \to \mu^{(i)}$ in eqn. (4.17), and then rescale $\mu^{(i)}$ and $M$. For example, we can take:

$$\left(\frac{\mu^{(i)}}{M}\right) = \left(\frac{\mu_o}{M_o}\right)\left(\frac{g(M)}{g(\mu)}\right)^{-\frac{\tilde{b}_1^{(i)}}{\tilde{b}_o^{(i)2}}} \qquad M = M_o g(M). \qquad (4.18)$$

In other words, it is necessary to integrate differently over different fields. This should not come as a surprise. If we had considered some sort of Pauli-Villars regulator fields as we did for our U(1) example, these would have come in complete $SU(5)$ multiplets, and the different components of the multiplets would be renormalized differently at low energies. Thus our Wilsonian cutoffs would be different for each gauge group. It is for this reason that we said in the introduction that we prefer the field-dependent cutoff language, since the rescaling in this case does not correspond to any simple redefinition of the unified coupling, $g(M)$. The rescaling in eqn. (4.18) is not unique. We will comment on this after we have considered non-perturbative effects in the next section.

### 5. Supersymmetric QCD

A somewhat more intricate example is provided by supersymmetric QCD with $N_f$ flavors of quarks and antiquarks. This theory is well-known to have flat directions, at the classical level, in which the gauge symmetry is completely or partially broken. Let us concentrate first on the case $N_f < N$. In these models there are flat directions with

$$Q = \bar{Q} = \begin{pmatrix} v_1 & 0 & \ldots & 0 \\ 0 & v_2 & \ldots & 0 \\ \ldots & \ldots & \ldots & \ldots \\ 0 & \ldots & \ldots & v_{n_f} \\ \ldots & \ldots & \ldots & \ldots \\ 0 & 0 & \ldots & 0 \end{pmatrix}. \qquad (5.1)$$

In these directions the symmetry is broken to $SU(N - N_f)$ if $N_f < N - 1$, and is completely broken for $N_f = N - 1$. The effective coupling in these directions is essentially the coupling of the full theory at the scale $v$.

If $N_f < N$, non-perturbative effects give rise to a superpotential. The form of this superpotential can be uniquely determined from the symmetries of the theory and the requirement of holomorphicity:

$$W = \Lambda^{\frac{3N-N_f}{N-N_f}} \det Q_o \bar{Q}_o^{-\frac{1}{N-N_f}} \qquad (5.2)$$

where the determinant is in flavor space. In stating that this result is exact, it is important that the chiral fields here must be understood as "bare," unrenomalized fields. In the case $N_f = N - 1$, for large $v$, the superpotential is generated by instantons; in the other cases, it is generated by gluino condensation in the unbroken group, $SU(N - N_f)$.[1]

In both cases, if one examines the detailed computations, one might expect complicated corrections in the coupling. Indeed, one might worry not only about non-holomorphic dependence on $g(M)$ but also on $g(v)$, or equivalently $\ln(|v|)$. Yet the general holomorphicity considerations we are invoking here show that this should not be the case, provided we work in terms of suitably defined couplings, or equivalently provided that we choose our cutoffs appropriately. In this section, we will show how this works for the first subleading corrections.

In the case $N_f < N - 1$ the non-perturbative superpotential arises as a result of gluino condensation in the $SU(N - N_f)$ theory; indeed, $W_{np}$ is proportional to $< \lambda\lambda >$.[1] Let us first examine the form of the gauge coupling function, along the lines described in the previous section. Work (for definiteness) in the momentum scheme, where we integrate out between some large scale $M$ and $\mu$. The massive vector supermultiplets will be taken to have mass $M_V = g(M_V)v$. Then, by the same logic as in the unified theory case,

$$\frac{8\pi^2}{g^2(\mu)} = \frac{8\pi^2}{g^2(M)} + \tilde{b}_o \ln(\mu/M) + \frac{\tilde{b}_1}{\tilde{b}_o} \ln\left(\frac{g(M)}{g(\mu)}\right)$$

$$+ (b_o - \tilde{b}_o) \ln(M_V/M) + \left(\frac{b_1}{b_o} - \frac{\tilde{b}_1}{\tilde{b}_o}\right) \ln\left(\frac{g(M)}{g(M_V)}\right). \qquad (5.3)$$

Here, as before, $b_o$ and $b_1$ are the $\beta$-function coefficients of the high energy theory; $\tilde{b}_o$ and $\tilde{b}_1$ are those of the broken phase. Explicitly:

$$b_o = 3N - N_f \qquad b_1 = 6N^2 - 2N_f N - 4N_f \frac{N^2-1}{2N}$$

$$\tilde{b}_o = 3(N - N_f) \qquad \tilde{b}_1 = 6(N - N_f)^2. \tag{5.4}$$

The vector mass in this expression is given by

$$M_V = gv = gv_o \sqrt{Z}^{-1} \tag{5.5}$$

where $v_o$ denotes the expectation value of the unrenormalized scalar field, and $\sqrt{Z}$ is the corresponding wave function renormalization factor:

$$M_V = gv_o \left(\frac{g(M)}{g(M_V)}\right)^{\frac{N^2-1}{2N}\frac{2}{b_o}}. \tag{5.6}$$

Using (5.6) and (5.4) we can rewrite eqn. (5.3) in terms of $v_o$:

$$\frac{8\pi^2}{g^2(\mu)} = \frac{8\pi^2}{g^2(M)} + \tilde{b}_o \ln(\mu/M) + \frac{\tilde{b}_1}{\tilde{b}_o} \ln\left(\frac{g(M)}{g(\mu)}\right)$$

$$+(b_o - \tilde{b}_o)\ln(v_o/M) + (b_o - \tilde{b}_o)\ln g(M). \tag{5.7}$$

In other words, only the expectation value of the bare scalar field appears here. All non-analyticity in $g$ then disappears if we write the coupling in terms of new scales, $\mu_o$ and $M_o$,

$$\mu = \mu_o \left(\frac{g(M)}{g(\mu)}\right)^{-\frac{\tilde{b}_1}{\tilde{b}_o^2}} \qquad M = M_o g(M)^{\frac{b_o - \tilde{b}_o}{b_o}}. \tag{5.8}$$

Clearly there is some freedom at this stage in the choice of rescaling; the reasons for the particular choice above will be clear shortly.

With this expression, it is now a simple matter to compute the superpotential through two loops. We can use a conventional renormalization group analysis to determine the form of the $<\lambda\lambda>$ condensate as a function both of $g(\mu)$ and $g(M)$. Choosing $\mu \ll M_V \ll M$

$$<\lambda\lambda> = \mu^3 \frac{g(M)^2}{g(\mu)^2} exp(-\frac{3 \times 8\pi^2}{\tilde{b}_o g^2(\mu)}) \tag{5.9}$$

(The dependence on $g(M)$ can be determined by writing a renormalization group equation for this object as a function of $M$, including the appropriate anomalous dimension.) Now use our earlier relation to write $\mu$ in terms of $\mu_o$. Using eqn. (5.8), one sees that

$$<\lambda\lambda> = \mu_o^3 exp\left(-\frac{3}{\tilde{b}_o}(\frac{8\pi^2}{g^2(M)} + (b_o - \tilde{b}_o)\ln(v_o/M_o) + \tilde{b}_o \ln(\mu_o/M_o))\right) \tag{5.10}$$

This yields precisely the non-perturbative superpotential of eqn. (5.2),

$$v_o^{-2N_f/(N-N_f)}(M_o e^{-\frac{8\pi^2}{b_o g^2(M)}})^{\frac{3b_o}{\tilde{b}_o}}. \tag{5.11}$$

Note that the result is, as expected, an analytic function of the bare fields and the couplings. More precisely, we have shown here that there are no terms involving $\ln(|v|^2)$ or $\ln(|g|^2)$. It is an elementary exercise to show that the result has the correct dependence on the phase of $v$ and the $\theta$-parameter.

The reason for our particular choice of rescalings in eqn. (5.8) is now clear. It is necessary to satisfy two holomorphy conditions: holomorphy of the gauge coupling and holomorphy of the gaugino condensate. Indeed, a major component of the analysis of SV is the holmorphy of the condensate. They prove this requirement by studying supersymmetry Ward identities. Alternatively, we can argue for it in the spirit of holomorphy as a function of couplings used in this paper. If we couple a chiral field, $S$, to $W_\alpha^2$, not only must the coupling function be analytic in $S$, but also any superpotential generated for $S$ (by gluino condensation, in particular) must be holomorphic. So the gluino condensate itself must be holomorphic. Indeed, as discussed in ref. 12, its precise dependence on $S$ can be determinend a priori from symmetries.

Let us now turn to the case $N_f = N - 1$, in which the superpotential is generated by instantons. The required instanton computation, including the requred one-loop determinant, has been performed by Cordes in ref. 16. Indeed, Cordes has considered the $g$-dependence of the calculation. Some features of this discussion, however, are slightly obscure. In particular, including only the one loop corrections to the instanton, one cannot determine whether the factors of $g$ which appear correspond to the coupling at the scale of the vector meson masses or at the scale of the cutoff. Here we give a slightly different description, in which we use dimensional analysis, improved by the renormalization group. We will see immediately that, if we work in terms of the *bare* fields, the superpotential is an analytic function both of the expectation values of the fields and of the "Wilsonian" gauge coupling.

The easiest quantity to compute with instantons is the mass of the light fermion. The result has dimensions of mass. The relevant scale in such computations, as stressed in the original work of 't Hooft,[17] is the vector boson mass, $g(M_V)v$. The result is then necessarily of the form

$$m_\chi = a g(M_V) v e^{-\frac{8\pi^2}{g^2(M_V)}}. \tag{5.12}$$

for some constant $a$. Substituting the explicit form of the two-loop $\beta$ function gives

$$m_\chi = a e^{-\frac{8\pi^2}{g^2(M)}} \frac{M^{2N+1}}{g(M)^{2N}} \frac{1}{v^{2N_f+2}} \left(\frac{g(M)}{g(M_V)}\right)^{\frac{4N_f}{b_o}\frac{N^2-1}{2N}}. \tag{5.13}$$

But from this we see that the superpotential is

$$W = \left(M_o e^{-\frac{8\pi^2}{b_o g(M)^2}}\right)^{2N+1} \frac{1}{det\bar{Q}Q\sqrt{Z}^{2N_f}}. \tag{5.14}$$

where

$$M_o = \frac{M}{g(M)^{2N/(2N+1)}} \tag{5.15}$$

Thus the bare superpotential satisfies all of the expected holomorphicity requirements.

This is a good point to return to the unified model, and consider the rescalings to be performed there. In that theory, there are two unbroken low energy groups, $SU(3)$ and $SU(2)$, with no matter fields. Clearly we want to require analyticity of the gauge coupling functions. What of the gluino condensates in the two groups? Consider, again, the coupling of the field $S$. Both the $SU(3)$ and $SU(2)$ condensates contribute to a superpotential for $S$, but the $SU(3)$ condensate is exponentially larger. In other words, the effects of the $SU(2)$ condensate are much smaller than any of the two-loop effects being considered in this paper. Indeed, the analysis of the $SU(2)$ condensate is complicated, for example, by the fact that higher dimension operators obtained by integrating out $M_{GUT}$ fields can induce an $SU(2)$ condensate independent of any pure $SU(2)$ dynamics. Thus at the level of our low order analysis, we should only impose the requirements of holomorphicity on the $SU(3)$ condensate. This yields a somewhat different set of rescalings then those given earlier. The rescalings of the $\mu^{(i)}$ are still different for the $SU(3)$ and $SU(2)$ groups.

## 6. A String Theory Application

We conclude by considering a problem in string theory in this language. Consider the question of the unification of couplings. It is well-known that in string theory the gauge couplings are unified at tree level (up to possible factors $k_a$ from the Kac-Moody algebras of the various gauge groups). The usual holomorphicity argument would then say that in perturbation theory, the only corrections to unification arise at one loop. If the string coupling is weak, any non-perturbative corrections will then be extremely small. However, there is good reason to think that if string theory describes nature, it is strongly coupled. Does this non-renormalization of the gauge couplings have any significance then?

In ref. 8, it is shown that in some cases, discrete gauge symmetries (which are expected to survive non-perturbatively) insure that any corrections to the gauge coupling function (and to the superpotential) are necessarily of the form $e^{-c \times 8\pi^2 S}$. In this reference, it is argued that even though string perturbation theory may not be valid, $S$ may – as observed in nature – be large, meaning that the effective gauge couplings are small. Potentially, then, the non-renormalization of the gauge couplings is a quite powerful statement about the full, non-perturbative string theory. Shifman and Vainshtein, on the other hand, have taught us that this non-renormalization is only true with a suitable definition of the coupling. One might worry that since this redefinition must be rediscovered at every order of perturbation theory (and beyond) that the non-renormalization is free of content. Here, however, the field-dependent cutoff language is very helpful. While in strong coupling, we do not expect the required rescaling of the cutoff to be computable, we also do not expect it to be exponentially large; indeed, we expect that it is of order some power of the coupling, i.e. of order one. This is of the same order as the uncertainties due to threshold effects; indeed, we expect thresholds to move by amounts of order one at strong coupling as well. So while we do not expect have complete control of these corrections, we do not expect them to be incredibly large.

This is, of course, both good and bad news. On the one hand, it means that string theory is more predictive than we might have expected. On the other hand, string theory is in danger of making the wrong prediction, at least if it produces a theory with MSSM particle content. It is conceivable that the cutoff's must be rescaled by factors of 100 or so, but this is an uncomfortable refuge.

In any case, this problem provides an example of a situation where the Shifman-Vainshtein program is potentially of more than academic interest: it provides a qualitative insight of quantitivative significance.

## Acknowledgements

We thank T. Banks, L. Dixon, R. Leigh, J. Polchinski and N. Seiberg for conversations, and J. Polchinski and N. Seiberg for reading an early version of this manuscript.